# Contraction or steady state? An analysis of credit risk management in Italy in the period 2008-2012

Stefano Olgiati MSc PhD, Prof Alessandro Danovi MSc
*University of Bergamo*

Presented at the New York School of Business – Copenhagen Business School International Risk Management Conference 2013: **Enduring Financial Stability – Contemporary Challenges for Financial Risk Management and Governance - Credit Risk and Tools for Financial Stability**, Copenhagen (DK) June 2013

## Abstract

Credit risk management in Italy is characterized, in the period June 2008 – June 2012, by frequent (frequency=0.5 cycles/year) and intense (peak amplitude: mean=39.2 b€; s.e.=2.83 b€) quarterly contractions and expansions around the mean (915.4 b€; s.e.=3.59 b€) of the nominal total credit used by non-financial corporations. Such frequent and intense fluctuations are frequently ascribed to exogenous Basel II procyclical effects on credit flow into the economy and, consequently, Basel III output-based point-in-time Credit/GDP countercyclical buffering advocated. We have tested the opposite null hypotheses that such variation is significantly correlated to actual default rates, and that such correlation is explained by fluctuations of credit supply around a steady state. We have found that, in the period June 2008 – June 2012 ($n$=17), linear regression of credit growth rates on default rates reveals a negative correlation of r=−.6903 with $R^2$=.4765, and that credit supply fluctuates steadily around the default rate with an Internal Steady State Parameter SSP=.00245 with $\chi^2$=37.47 ($v$=16, P<.005). We conclude that fluctuations of the total credit used by non-financial corporations are exhaustively explained by variation of the independent variable "default rate", and that credit variation fluctuates around a steady state. We conclude that credit risk management in Italy has been effective in parameterizing credit supply variation to default rates within the Basel II operating framework. Basel III prospective countercyclical point-in-time output buffers based on filtered Credit/GDP ratios and dynamic provisioning proposals should take into account this underlying steady state statistical pattern.

# 1. Background

Credit risk management has become one of the most relevant topics both for financial institutions and for scholars. Credit risk models have evolved from subjective analysis to accounting-based credit-scoring systems and measures of credit risk and risk concentration (Altman and Saunders, 1998) and their effects on capital allocation and shareholders' value in banking assessed (Resti and Sironi 2012).

The European Commission with the Credit Risk Directives (CRD I, II and III) and Banking Authorities with Basel Accords on minimum capital requirements and countercyclical buffers (Basel II and III) are still carrying out a long process of formalization of credit risk management methods and guidelines in order to diffuse a culture of common rules at the continental level.

Monitoring, data collecting and analysis of economic and financial cyclicality is coordinated in the EU by Eurostat, with cyclical indicators[1] such as the Business Climate Indicator (BCI), the OECD Composite Leading Indicators (CLI), the Ifo Economic Climate Indicator, the DZ Euroland, the IARC, IESR and E-Coin published quarterly by Eurostatistics.

Eurostat has developed and implemented a set of guidelines for the statistical analysis of cyclical fluctuations (2003) and modern statistical tools (Sigma 2009) to which we will refer in full[2].

As far as banks' regulatory capital is concerned, procyclicality, and the potential effects of capital requirements standards on the flow of credit into the economy, have been addressed by the Basel II Committee and Italy's Central Bank (Banca d'Italia)[3] which recommended to use long term data horizons to estimate probabilities of default ($PD$)[4], to introduce a downturn loss-given-default ($LGD$) estimate[5] and to introduce expected long-run loss rates ($EL$) in AIRB methods[6]. Basel II accords require own estimates of $PD$ and $LGD$ to be no less than the long-run default-weighted average loss rate given default calculated based on the average economic loss of all observed defaults within the data source for that type of facility[7]. Coherently, the introduction of point-in-time output buffers based on a Hodrick-Prescott filter of the macroeconomic Credit-to-GDP gap[8] to reduce procyclicality during periods of excessive credit growth and promote countercyclical dampening during periods of contraction is among the main goals of the ongoing Basel III reform[9].

---

[1] Eurostatistics 12/2012: 9-14
[2] Eurostat (2003); **3.2**
[3] Banca d'Italia (2006), *Nuove disposizioni di vigilanza prudenziale per le banche - Circolare n. 263 del 27 dicembre 2006*
[4] See BCBS 2006, sub-sections 472, 502, 503, 504.
[5] See BCBS 2006, sub-section 468
[6] See BCBS 2006, sub-section 367 and Table 6 page 236
[7] See BCBS 2006, sub-section 468
[8] See BCBS 2010a, pages 8-14

Specifically, Italy is characterized by the enduring effects of the 2007-09 financial crisis in terms of actual and prospective negative GDP growth (-2.4% in 2012; -0.2% in 2013), growing sovereign Debt (≅2,000 b€) and a growing Debt/GDP ratio (≅1.25) ratio[10] In the period June 2008 – June 2012, the volume of outstanding loan facilities is characterized by frequent (frequency=0.5 cycles/year) and intense (peak amplitude: mean=39.2 b€; s.e.=2.83 b€) quarterly cyclical fluctuations[11] in the minima to maxima[12] interval around the mean (915.4 b€; s.e.=3.59 b€) of the nominal total credit used by non-financial corporations[13].

The conflicting effects of cyclicality on the tradeoff between stability and timeliness in predicting probabilities of default and recovery rates have been analyzed by Altman, Brady, Sironi and Resti (2005), who observe that banks tend to react to short-term evidence therefore regulation should encourage the use of long-term average rates in AIRB systems. In Italy linear long-term predictions due to the frequent cyclical waveform fluctuation are statistically significant ($\hat{y} \cong y$ and $d\hat{y}/dx \cong dy/dx$) only every 8 quarters (4 phases, 2 years).

Altman and Rijiken (2005) observe that agencies delay the timing of through the cycle rating migration estimates by 0.56 years at the downside and 0.79 years at the upside. This signifies that in Italy, with a phase period of 0.5 years, as we will see, in a period economic downturn, agency ratings are systematically one phase late through the cycle.

Jarrow et al. (1997) provide a discrete time-homogeneous Markov chain transition matrix for the term structure of credit risk spreads which assumes a time step of one year. In Italy in the period 2008-2012 this time step corresponds to two phases of the cycle (1 year), rendering the assumption of time-homogeneity during such time step not statistically acceptable.

Gordy and Howells (2004) observe that credit risk adjusted portfolio management is based on time-homogeneous Markov transition processes, which are based on *ex-ante* probabilities of default which register all *expected* variation in the rating variables and register all *ex-post* variation as *unexpected*.

---

[9] See BCBS 2010a, page 1

[10] MINEF, Documento di Economia e Finanza 2012, II: Documento di analisi e tendenze di finanza pubblica

[11] If a period is the duration of 1 cycle, the frequency is the number of cycles per period. The amplitude is the minima and maxima absolute values of the cycle. In our case: period=2 years, then frequency=1/2=0.5 cycles/year. In physical notation, to which we refer in this paper, a cycle has 4 phases: $dy/dx>0$ $d^2y/dx^2>0$, $dy/dx>0$ $d^2y/dx^2<0$, $dy/dx<0$ $d^2y/dx^2<0$, $dy/dx<0$ $d^2y/dx^2>0$, 1 minimum $dy/dx=0$ $d^2y/dx^2>0$ and 1 maximum $dy/dx=0$ $d^2y/dx^2<0$. The phase period is equal to the cycle period/4.

[12] In a discrete distribution a maximum is determined when $y(t)>y(t-1)$ and $y(t)>y(t+1)$, a minimum when $y(t)<y(t-1)$ and $y(t)<y(t+1)$ and a steady state when $y(t)=y(t-1)$ and/or $y(t)=y(t+1)$.

[13] Italy Central Bank, Statistical Bulletin  III-2012, Information on Customer and Risk, Default Rates For Loan Facilities And Borrowers, TDB30486: Quarterly default rates for loan facilities - Distribution By Customer Sector Of Economic Activity And Total Credit Used: Non-financial Corporations - Reporting Institutions: Banks, Financial Companies And Other Institutions Reporting To The Ccr

Frequent cyclicality would systematically alter the ratio between unexpected and expected variation. Repullo et al. (2008, 2009, 2011) observe that higher buffers in expansions are insufficient to prevent a significant contraction in the supply of credit at the arrival of a recession, which in Italy has occurred in the period 2008-2012 every year.

Sironi and Resti (2012) observe that a modification of the current IFRS 39 concept of incurred loss with a principle of fair value and amortized cost could further increase the procyclicality of banks' credit policies. In Italy, 0.5-year phases render misleading through-the-cycle quarterly and half year estimates of fair values.

## 2. Research Questions and Methods

In this paper we have asked two research questions:

**Q1** – is there a statistically significant linear relationship linking credit output fluctuations to default rates in the period June 2008 – June 2012? We argue that if such a linear relationship does exist or, in other words, if variation in credit supply is satisfactorily explained by independent variation in the default rates then, given the *a priori* postulate that exogenous macro-conditions, such as the business cycle, do affect default rates, and Basel II minimum capital requirements do have procyclical effects, then such exogenous effects are satisfactorily *transformed* by the relationship between credit and default rates, as it should be according to operating Basel II Accords;

**Q2** – given that Q1 linear relationship does exist, can we formulate a null hypothesis regarding the causes of such relationship which can be statistically analyzed and tested? In particular we will test the hypothesis that credit supply variation systematically converges towards a steady state, i.e. credit supply is systematically increased or decreased in order to achieve credit steady state at a definite level parameterized by a steady state parameter.

We have analyzed Italy's Central Bank Statistical Bulletin's quarterly default rates for loan facilities (credit used) in the period March 1996 – June 2012: Information on customer and risk, default rates for loan facilities and borrowers (TDB30486); Quarterly default rates for loan facilities; Distribution by customer sector of economic activity and total credit used: Non-financial corporations; Reporting institutions: Banks, financial companies and other institutions reporting to the Central Credit Registrar.

Coherently, we have defined:

*ABD* = Adjusted bad debts refer to the total loan exposure of borrowers who, for the first time in the reference quarter, meet one of the total loans outstanding when a borrower is reported to the central credit register: a) as a bad debt by the only bank that disbursed credit; b) as a bad debt by one bank and as having an overshoot by the only other bank exposed; c) as a bad debt by one bank and the amount of the bad debt is at least 70% of its exposure towards the banking system or as having overshoots equal to or more than 10% of its total loans outstanding; d) as a bad debt by at least two banks for amounts equal to or more than 10% of its total loans outstanding;

*TCU* = the amount of <u>total credit used</u> by all the borrowers covered by the central credit register and not classified as adjusted bad debtors at the end of the previous quarter. The *TCU* does not include the credits that, in the given quarter, have been transferred to institutions not reporting to the central credit register;

*d* = The <u>default rate of loan facilities in a given quarter</u> is represented by the ratio between the amount of total credit used by borrowers who become adjusted bad debtors (*ABD*) during the quarter in question and the amount of credit used by all the borrowers covered by the central credit register and not classified as adjusted bad debtors at the end of the previous quarter (*TCU*);

*L* = Loans refer to <u>loans disbursed by banks to non-banks calculated at face value</u> (until September 2008 at book value) gross of adjustment items and net of repayments. The aggregate includes mortgage loans, current account overdrafts, loans secured by pledge of salaries, credit card advances, discounting of annuities, personal loans, leasing (from December 2008 according to the ias17 definition), factoring, other financial investments (e.g. commercial paper, bill portfolio, pledge loans, loans granted from funds administered for third parties), bad debts and unpaid and protested own bills. the aggregate is net of repurchase agreements and, since December 2008, net of stock exchange repos and gross of correspondent current accounts. performing loans.

We have analyzed data with a simplified Bayesian biostatistical technique called "Retrospective Forecasting" utilized by Shaman and Karspeck (2012a , 2012b) to predict flu epidemics in New York City on the basis of fluctuating outcomes. The technique assumes retrospectively perfect knowledge of future parameters and the posterior parameters and other state variables are reset to an initial distribution before commencing each reiterative forecast form the present into the past (Backward Calculation)[14] which, as we will see, will determine, in our case, the Internal Steady State Parameter ς (*little sigma*) of the system.

An epidemic reaches a steady state when:

$$H^I : 1 = \left(1 + f_{0,1}\right)\left(1 - d_{0,1}\right) + \frac{\left(1 + f_{1,2}\right)\left(1 - d_{1,2}\right)}{1 + \varsigma} + \ldots + \frac{\left(1 + f_{n-1,n}\right)\left(1 - d_{n-1,n}\right)}{\left(1 + \varsigma\right)^n}$$

$$s = \frac{1}{1 + \varsigma}$$

$$1 = \left(1 + f_{0,1}\right)\left(1 - d_{0,1}\right) + \left(1 + f_{1,2}\right)\left(1 - d_{1,2}\right)s + \ldots + \left(1 + f_{n-1,n}\right)\left(1 - d_{n-1,n}\right)s^n$$

$$\varsigma = \left(1/s\right) - 1 = 0$$

where we have defined the variable *f* as:

[14] See also Backward Calculation in Eurostat (2003); **3.2**

$f$ = The <u>credit growth rate</u> determined as $f_{t,t+1} \simeq \left[\dfrac{L_{t+\delta,t+1}}{TCU_{t+\delta}\left(1-d_{t+\delta,t+1}\right)}\right]_{\delta \to 0^+}$ .

The credit growth rate is asymptotically equal to the amount of loan facilities disbursed in period $t,t+n$ divided by the total credit used at the beginning $t$ of the period which will survive ($1$-$d$) to the end $t+n$ of such period. The variable $f$ assumes that the credit risk manager possesses perfect information regarding $d$ and therefore will not grant new loans or additional loans $L$ to borrowers which he knows will default in the same period;

$\delta$ = The <u>sliding time parameter</u> $\delta$ accounts for the fact that the nearer the credit risk manager gets retrospectively to time $t+n$, the more perfect information becomes and thus the retrospective forecast. In the quarterly data analyzed we have assumed $\delta$=0, i.e. the credit risk manager has at time $t$ perfect information regarding default rates and credit growth rates regarding the period $t,t+n$.

We have assumed an LGD=1; the hypothesis is reasonable in the framework of the analysis since recovery rates affect, at time of recovery, the credit supply to the economy, and such effect will be "seen" in the total credit used and in the credit growth rate.

We can now formulate the null hypotheses that:

Q1: $H_0^{Q1}: f_{t,t+n} = \beta_1 + \beta_2 d_{t,t+n} + \varepsilon_{t,t+n}$ : credit supply growth is a dependent variable explained by a $\beta_2 < 0$ negative linear relationship with the independent default rate;

Q2: $H_0^{Q2}: f = \dfrac{d}{1-d}$ : credit supply growth or decline rate is explained by odds of default, i.e. the ratio between the default rate and the survival rate. In alternative, we will test the hypothesis that credit supply growth or decline rate is not explained by the odds of default alone and is sensitive to exogenous cyclical positive or negative factors, which we will define as the Steady State Parameter $\varsigma$ of the system. There follows that $H_1^{Q2}: f = \dfrac{d+\varsigma}{1-d}$

We have utilized Mathematica 8 and Statistical-Graphical Integration with Mac OS X Datagraph 3.1.

## 3. Findings

We have divided the Credit Supply Growth Rates ($f$) and the Default Rates ($d$) of the period March 1996 – June 2012 into 2 sub-periods. The first period from March 1996 to June 2008 excluded, and the second period from June 2008 included to June 2012.

Linear regression of credit supply rates on default rates by Ordinary Least Squares (OLS) from March 1996 to June 2008 ($n$=49) reveals a not significant r=.20817; $R^2$ =.043336 and $s$ for

residual=.024809. Steady state null hypotheses $H_0$ and $H_1$ testing of the observed credit growth rates vs. the expected rates reveals a steady state parameter of Ç=.020584 (2.1%) with a $s$ for residual=.024976 and $\chi^2$= 106.66 with $v$=48 (Exhibit 1 – Functions not shown).

Linear regression of credit supply rates on default rates by Ordinary Least Squares (OLS) from June 2008 to June 2012 ($n$=17) reveals a r=−.69032; $R^2$ =.47654 and $s$ for residual=.0087185. Steady state null hypotheses $H_0$ and $H_1$ testing of the observed credit growth rates vs. the expected rates reveals a steady state parameter of Ç=.00245 (0.2%) with a $s$ for residual=.013152 and $\chi^2$= 37.47 ($v$=16, P<.005) (Exhibit 1).

In synthesis, from June 2008 to June 2012 ($n$=17), we accept the hypothesis that credit growth rates are negatively correlated to default rates but appear to be significantly fluctuating around the Steady State Function.

The heuristic path of adjustment of credit growth rate in the period June 2008-June 2012 to the Steady State Function Ç=.00245 is shown in Exhibit 2

**Exhibit 1**

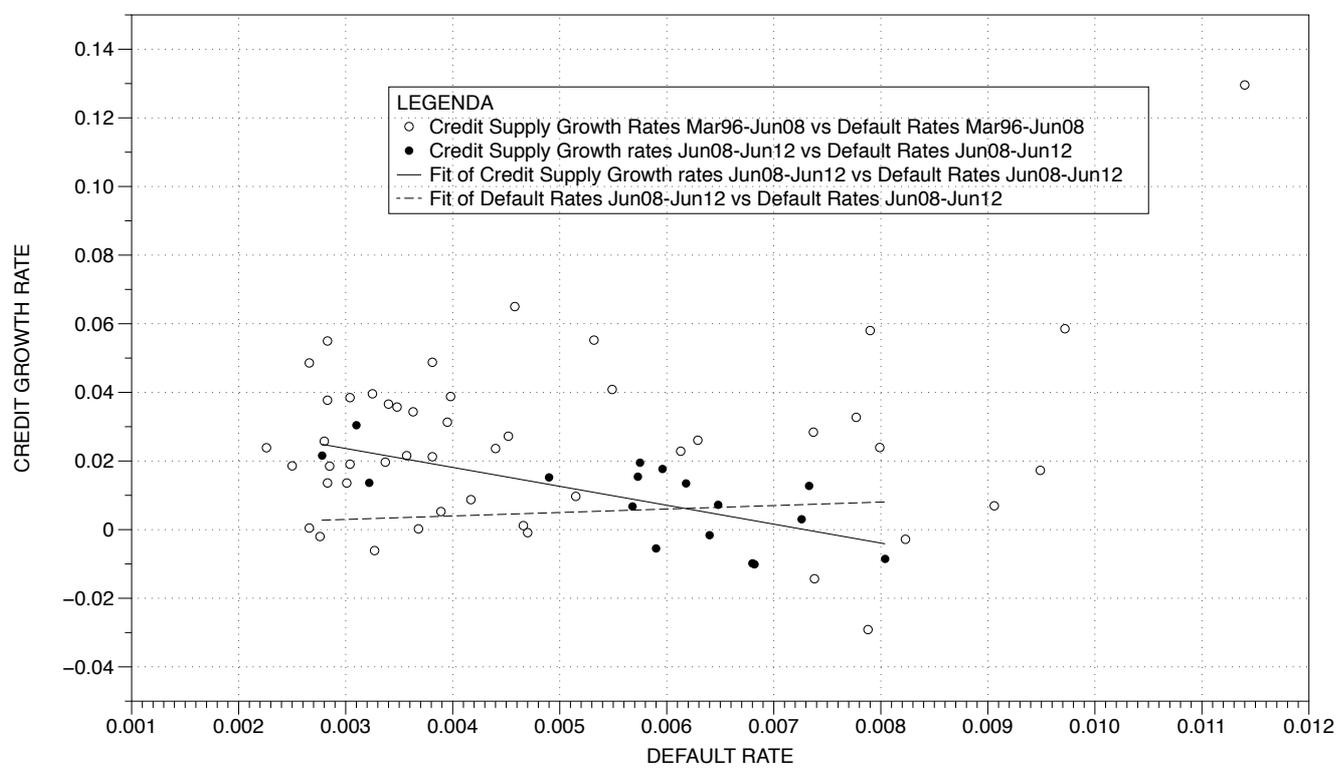

**Sources**:
Banca d'Italia TDB30486, ISTAT, Ministero dell'Economia e delle Finanze
**Statistics**:
Mathematica 8 and Mac OS X Datagraph 3.1

# Exhibit 2: Credit Supply Growth Rate fluctuations around the exogenous Steady State Parameter Ϛ in the period June 2008 – June 2012

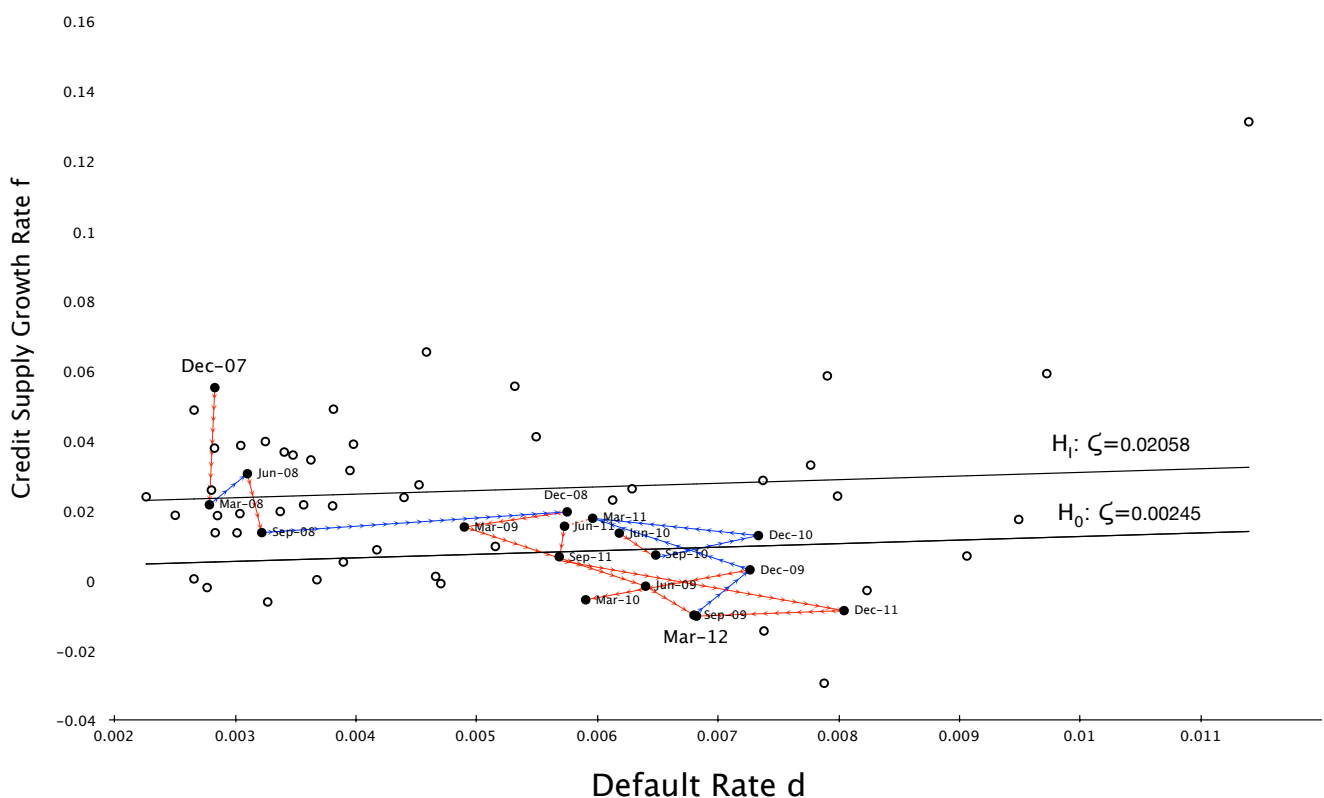

# Exhibit 3: Parameters of Ordinary Least Squares (OLS) Linear Regression and Steady State Function (SSF)

|  | [March 1996 – June 2008) | | [June 2008 – March 2012] | |
|---|---|---|---|---|
|  | OLS | SSF | OLS | SSF |
| N | 49 | **49** | 17 | **17** |
| Ϛ |  | **0.020584** |  | **0.00245** |
| Intercept | 0.0144 |  | 0.040187 |  |
| σ Intercept | 0.0086 |  | 0.0092139 |  |
| X Intercept | -0.0063 |  | 0.0072893 |  |
| Slope | 2.2931 |  | -5.5131 |  |
| σ Slope | 1.6061 |  | 1.5443 |  |
| Correlation | 0.20817 |  | -0.69032 |  |
| R2 | 0.04334 |  | 0.47654 |  |
| σ | 0.024549 | **0.024715** | 0.0084582 | **0.01276** |
| s for residual | 0.024809 | **0.024976** | 0.0087185 | **0.013152** |
| $\chi^2$ | | **106.66** | | **37.47** |

# 4. Interpretation

Credit risk management in Italy is characterized, in the period June 2008 – June 2012, by frequent (frequency=0.5 cycles/year) and intense (peak amplitude: mean=39.2 b€; s.e.=2.83 b€) quarterly contractions and expansions around the mean (915.4 b€; s.e.=3.59 b€) of the nominal total credit used by non-financial corporations. Such frequent and intense fluctuations are frequently ascribed to exogenous Basel II procyclical effects on credit flow into the economy and, consequently, Basel III output-based point-in-time Credit/GDP countercyclical buffering advocated. We have tested the opposite null hypotheses that such variation is significantly correlated to actual default rates, and that such correlation is explained by fluctuations of credit supply around a steady state. We have found that, in the period June 2008 – June 2012 ($n$=17), linear regression of credit growth rates on default rates reveals a negative correlation of r=−.6903 with $R^2$=.4765, and that credit supply fluctuates steadily around the default rate with a Steady State Parameter SSP=.00245 with $\chi^2$=37.47 ($v$=16, P<.005). We conclude that fluctuations of the total credit used by non-financial corporations are exhaustively explained by variation of the independent variable "default rate", and that credit variation fluctuates around a steady state. We conclude that credit risk management in Italy has been effective in parameterizing credit supply variation to default rates within the Basel II operating framework. Basel III prospective countercyclical point-in-time output buffers based on filtered Credit/GDP ratios and dynamic provisioning proposals should take into account this underlying steady state statistical pattern.

# 5. Limits

As Gordy (2003) observes credit risk is idiosyncratic to the obligor, and what we define a cycle is really a composite of a multiplicity of cycles tied to location, period and sector. Therefore this model suffers form the same limits as the Credit/GDP countercyclical buffers, i.e. a single-factor model cannot capture any clustering of default rates due to dishomogeneous sensitivity to smaller-scale components of the macro cycle.

# 6. References

[1] Altman E.I., Saunders A. (1996), "Credit risk measurement: Developments over the last 20 years", *Journal of Banking & Finance;* **21**: 1721-1742

[2] Altman E.I and Rijken H.A. (2005), "The effects of rating through the cycle on rating stability, rating timeliness and default prediction performance", http://archive.nyu.edu/bitstream/2451/26405/2/FIN-05-004.pdf, March (accessed 12/23/12)

[3] Altman E.I., Brady B., Resti A., Sironi A. (2005), "The Link between Default and Recovery Rates: Theory, Empirical Evidence and Implications", *Journal of Business*; **78**: 6

[4] Angelini P., Gerali A. (2012), "Banks' reactions to Basel-III", *Banca d'Italia Working Papers*, July; Number 876: 5-26

[5] Banca d'Italia (2006), *Nuove disposizioni di vigilanza prudenziale per le banche - Circolare n. 263 del 27 dicembre 2006*

[6] Banca d'Italia, *Statistical Bulletin On-line, Information on Customer and Risk*, http://bip.bancaditalia.it/4972unix/homebipentry.htm?dadove=corr&lang=eng (accessed 1/1/13)

[7] Basel Committee on Banking Supervision (2006), "International Convergence of Capital Measurement and Capital Standards. A Revised Framework Comprehensive Version", Bank for International Settlements, Basel, June

[8] Basel Committee on Banking Supervision (2010a), "Guidance for national authorities operating the countercyclical capital buffer", Bank for International Settlements, Basel, December

[9] Basel Committee on Banking Supervision (2010b), "Basel III: A global regulatory framework for more resilient banks and banking systems", Bank for International Settlements, Basel, December (rev June 2011)

[10] Berger A.N., Udell G.F. (2002), "The Institutional Memory Hypothesis and the Procyclicality of Bank Lending Behavior", *BIS Working Papers,* Board of Governors of the Federal Reserve System, December

[11] Borio C., Furfine, C., Lowe P (2001): "Procyclicality of the financial system and financial stability: issues and policy options", *BIS papers*; **1**: 1-57

[12] Catarineu-Rabell E., Jackson P., and Tsomocos D. (2003), "Procyclicality and the new Basel Accord—Banks' choice of loan rating system," *Economic Theory*, No 26, 2005

[13] Elliott, D. (2009), "Quantifying the effects on lending of increased capital requirements", Pew Financial Reform Project Briefing Paper No. 7.

[14] Eurostat (2003), *Statistical analysis of cyclical fluctuations*, Office for Official Publications of the European Communities, Luxembourg

[15] Eurostat (2012), *Eurostatistics Statistical books  - Data for short-term economic analysis*, Office for Official Publications of the European Communities, Luxembourg

[16] Gordy M.B. (2003), "A risk-factor model foundation for ratings-based bank capital rules", *Journal of Financial Intermediation*; **12**:199–232

[17] Gordy M.B. and Howells B. (2004),  "Procyclicality in Basel II: Can We Treat the Disease Without Killing the Patient?", Bank of International Settlements, Basel, May

[18] Jarrow R.A., Lando D., Turnbull S.M. (1997), "A Markov Model for the Term Structure of Credit Risk Spreads", *The Review of Financial Studies,* Summer 1997; Vol. 10, No. 2: pp. 481–523

[19] Kashyap A.K., Stein J.C. (2004), "Cyclical implications of the Basel II capital standards", Federal Reserve Bank of Chicago, Economic Perspectives, 1Q/2004

[20] Keeton W.R. (1999), "Does Faster Loan Growth Lead to Higher Loan Losses?", *Economic Review  - Second Quarter 1999*, *Federal Reserve Bank Of Kansas City;* **II**: 57-75

[21] Matricciani E. (2009), *Lezioni di probabilità e processi aleatori,* 1[st] Ed., Leonardo, Bologna

[22] Ministero dell'Economia e delle Finanze, *Documento di Economia e finanza 2012*, http://www.tesoro.it (accessed 1/1/13)

[23] Owen F., Jones R. (1994), *Statistics*, 4[th] Ed., Pitman, London

[24] Repullo R., J. Suarez (2008), "The Procyclical Effects of Basel II", CEPR Discussion Paper, No. 6862

[25] Repullo R., Saurina J., Trucharte C. (2009), "Mitigating the Procyclicality of Basel II", *Macroeconomic Stability and Financial Regulation: Key Issues for the G20*, Centre for Economic Policy Research (CEPR): 105-12

[26] Repullo R., Saurina J. (2011), "The countercyclical capital buffer of basel III a critical assessment", *Banco de España*, March

[27] Resti A., Sironi A. (2012), *Risk management and shareholders' value in banking*, Wiley, New York: 623-28

[28] Ross S. (2002), *A first course in probability*, 6[th] Ed., Prentice-Hall, NJ

[29] Saurina, J.,Trucharte C. (2007), "An ASSFssment of Basel II Procyclicality in Mortgage Portfolios," *Journal of Financial Services Research*; **32**: 81-101.

[30] Shaman J. and Karspeck A. (2012a), "Forecasting seasonal outbreaks of influenza", *PNAS*; published ahead of print November 26, 2012, doi:10.1073/pnas.1208772109

[31] Shaman J. and Karspeck A. (2012b), "Forecasting seasonal outbreaks of influenza – Supporting Information", *PNAS*; doi:10.1073/pnas.1208772109

[32] Sigma (2009), "Modern tools for business cycle analysis", *The bulletin of European statistics*, Office for Official Publications of the European Communities, Luxembourg

[33] UK FSA (2009), "Variable Scalar Approaches to Estimating Through the cycle PDs", http://www.fsa.gov.uk/pubs/international/variable_scalars.pdf, February ((acceSSFd 12/23/12)

[34] Woolfson M.A., Woolfson M.S. (2007), *Mathematics for physics*, 1[st] Ed., Oxford University Press, Oxford